\documentclass[apl,twocolumn,floats]{revtex4}%
\usepackage{graphicx,epsfig}% Include figure files

\begin{document}

%\preprint{APS/123-QED}

\title{Tunneling and nonlinear transport in a vertically coupled GaAs/AlGaAs double quantum wire system}

\author{E. Bielejec}
 \email{esbiele@sandia.gov}
\author{J. A. Seamons}
\author{J. L. Reno}
\author{M. P. Lilly}
\affiliation{Sandia National Laboratories, Albuquerque, NM 87185}

\date{\today}

\begin{abstract}
We report low-dimensional tunneling in an independently contacted
vertically coupled quantum wire system.  This nanostructure is
fabricated in a high quality GaAs/AlGaAs parallel double quantum
well heterostructure.  Using a novel flip chip technique to align
top and bottom split gates to form low-dimensional constrictions
in each of the independently contacted quantum wells we explicitly
control the subband occupation of the individual wires. In
addition to the expected 2D-2D tunneling results, we have found
additional tunneling features that are related to the 1D quantum
wires.
\end{abstract}

\pacs{Valid PACS appear here}  % PACS, the Physics and Astronomy
                               % Classification Scheme.
%\keywords{Suggested keywords} %Use showkeys class option if keyword
                               %display desired
\maketitle

Coupled nanostructures show promise in leading to new
understanding of non-fermi liquid physics, many-body effects and
electron-electron interactions
\cite{Zulicke2002,Eugster1991,Eugster1994,Smoliner1996,Ploner2000,Auslaender2002,Tserkovnyak2002,Bird2003}.
The vertically coupled double quantum wire system described in
this letter is a simple realization of such a coupled
nanostructure. Proposals of coupled quantum wire devices that
utilize tunneling and spin-orbit coupling to make a spin filter
\cite{Governale2002} and coherent tunneling oscillations to form a
qubit \cite{Bertoni2000} rely on exquisite control of the wire
density, number of subbands occupied and coupling both between the
wires and to the macroscopic environment.

In this letter we report tunneling measurements between vertically
coupled double quantum wires fabricated using split gates on both
sides of a double quantum well GaAs/AlGaAs heterostructure.  While
semiconductor based nanoelectronics promise a flexible platform,
care must be taken that implementation of the quantum wires and
the control over their subband occupation does not also alter the
properties of the wires under investigation.  With this issue in
mind, we devise a tunneling geometry where the quantum wires are
separated from the gates that are necessary for independent
contacts.  The price for this isolation is a 2D-2D tunneling
component.  We present tunneling results where 2D-2D resonances
and additional structures are visible. The origin of the
additional features and their relationship to 1D physics is
investigated.

For the samples reported here, 18 nm wide GaAs quantum wells are
separated with an AlGaAs barrier. For sample A the barrier is 7.5
nm and for sample B the barrier is 10 nm. The top and bottom
quantum wells in sample B have individual layer densities of
$1.16$ and $1.96 \times 10^{11}$ cm$^{-2}$ respectively and a
combined mobility of $0.91 \times 10^{6}$ cm$^{2}$/Vs; sample A is
similar. The heterostructure is thinned to approximately 0.4
$\mu$m using an epoxy-bond-and-stop-etch (EBASE) process
\cite{EBASE1996}. This allows for top and bottom split gates
$\sim$ 150 nm from the top and bottom quantum wells, aligned
laterally with sub-0.1 $\mu$m resolution. The advantages of this
device structure are three-fold. First, by making use of molecular
beam epitaxy (MBE) growth of the tunneling barrier we have a rigid
potential barrier between the layers. Second, the close proximity
of the top and bottom split gates to the electron layers leads to
a well-defined confinement potential for the 1D wires. Third,
independent contact to individual electron layers in combination
with the top and bottom split gates allows for the independent
formation and control of the number of occupied subbands in both
the top and bottom quantum wires.

\begin{figure}[t]
\centerline{\epsfig{file=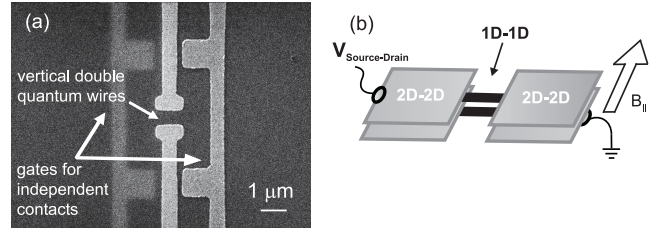,width=8.5cm}} \caption{(a)
Scanning electron microscope image of a coupled quantum wire
device. (b) Schematic diagram of the resonant tunneling
measurement for this device.  Both 1D-1D and 2D-2D tunneling can
occur in this device. } \label{Figure 1}
\end{figure}

In Fig. 1(a), a top view scanning electron micrograph of a coupled
quantum wire device is shown.  The dark areas are the GaAs/AlGaAs
heterostructure described above.  The active region of the device
consists of six TiAu gates. The four gates in the center form
pairs of split gates defining the quantum wires.  Only the two
gates on the top surface are visible due to the accurate alignment
of the top and bottom gates. The quantum wires are formed by
electrostatic confinement in the 0.5 $\mu$m wide and 1.0 $\mu$m
long gap between the split gates. The remaining two gates,
spanning the entire field of view, are used to independently
contact the individual electron layers. The lateral separation
between the quantum wire and these gates is chosen to be $\sim$ 10
times greater than the depth of the quantum well to ensure a
uniform density of the quantum wires. Low frequency measurements
of the parallel transport and tunneling conductance are made using
a standard ac method with a constant excitation voltage of 100
$\mu$V at 13 to 143 Hz; dc IV's are used as a consistency check.
All measurements were taken at base temperature of a dilution
refrigerator with an approximate electron temperature of 50 mK.

\begin{figure}[t]
\centerline{\epsfig{file=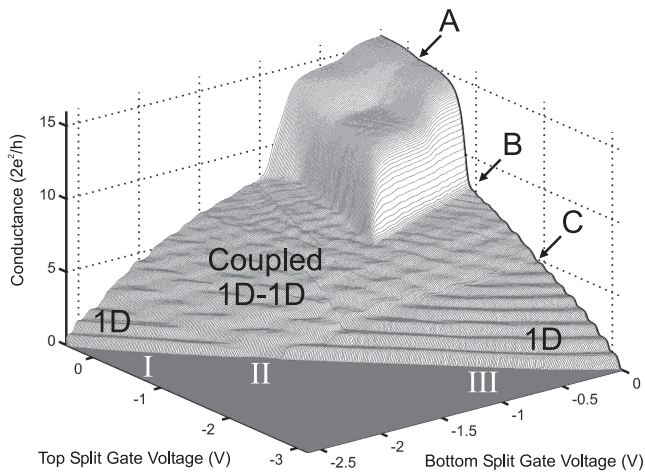,width=8.5cm}} \caption{
Two-terminal parallel transport with respect to the top and bottom
split gate voltage for sample A at $T =$ 50 mK.  The split gate
voltages explicitly control the occupation of the top and bottom
quantum wires. The left (I) and right (III) of the figure are
regions where only one of the quantum wires are occupied as
indicated by uniform conductance steps (quantized in 2$e^{2}/h$
after subtracting series contact resistance). The center region
(II) is where both wires are occupied.} \label{Figure 2}
\end{figure}

In Fig. 2 we plot the two-terminal linear conductance of the two
layers in parallel as a function of the top and bottom split gate
voltage for sample A. This figure shows the transition from 2D to
1D and finally to zero conductance as negative top and bottom
split gate voltage is applied to the system.  Consider the case
where the bottom split gate voltage is fixed at zero and we vary
the top split gate voltage.  For top split gate voltages up to
point A (Fig. 2 bold line) we have two 2D electron layers present.
As we apply a larger negative top split gate voltage (-0.33 to
-0.81 V, A to B) a wide 1D channel is formed in the top electron
layer in parallel with the lower 2D layer. For larger negative top
split gate voltage (-0.81 to -1.76 V, B to C) 1D constrictions
form in each of the layers, although the top constriction is much
narrower than the bottom constriction. By a top split gate voltage
of -1.76 V (C) the top layer is completely completely depleted and
conductance steps are observed in the bottom wire. The presence of
quantized conductance in region III is an indication of the
occupation of the bottom wire only.  Similarly, in region I only
the top wire is occupied.  In the center region (II) of the figure
we observe a more complicated pattern when both of the wires are
occupied and contribute to the conductance.  The complicated
crossing pattern provides a means to map out the subband
occupation and illustrates the control we have over the states of
the individual wires.

A schematic view of the resonant tunneling geometry is shown in
Fig. 1(b).  Using this tunneling geometry we expect a combined
1D-1D and 2D-2D tunneling signal.  The 1D-1D results from the
overlap of the quantum wires themselves and the 2D-2D results from
the small 2D areas on either side of the quantum wires.  Resonant
tunneling occurs when an electron in one wire tunnels to the other
wire conserving both energy and momentum \cite{Zulicke2002}.

In Fig. 3(a) we show the 2D-2D ac tunneling conductance for
several values of magnetic field, $H_{||}$, applied parallel to
the plane of the two-dimensional electron system and perpendicular
to the tunneling current (see Fig. 1(b)).  2D-2D tunneling occurs
when the top and bottom quantum wire split gates are grounded,
while maintaining independent contact to the individual layers
(see Fig, 1(a)).  For $H_{||} =$ 0 a sharp resonance occurs at
$\sim$ 2.8 mV.  We observe a decrease in the tunneling conductance
and a splitting of the resonance with increasing magnetic field.
This decrease and splitting has been previously observed
\cite{Eisenstein1991,Smoliner1989,Smoliner1995} and is attributed
to the magnetic field shifting the dispersion curves in k-space
which reduces the phase space for tunneling and leads to a complex
dependence on both the magnetic field and applied
V$_{Source-Drain}$ (V$_{SD}$). The observed 2D-2D tunneling at
$\sim$ 2.8 mV is in good agreement with the measured densities of
the two layers via Shubnikov-deHaas oscillations and gate
depletion studies. Due to the presence of small 2D areas on either
side of the quantum wires, the 2D-2D tunneling features described
here are visible to some extent in all our tunneling results.

\begin{figure*}[t]
\epsfig{file=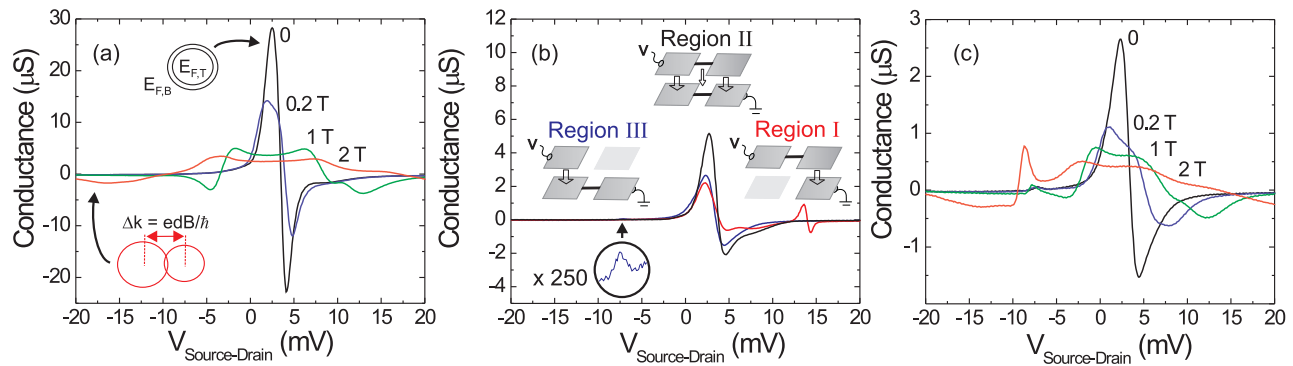,width=17cm} \caption{(a) 2D-2D tunneling
for H$_{||}$ ranging from 0 to 2T as indicated on the figure for
sample B at $T =$ 50 mK.  We indicate the top and bottom Fermi
circle positions for the 0 and 2T data. (b) Tunneling conductance
from regions I, II and III as labelled in Fig. 2. Additional side
peaks appear in both regions I and III.  For region III the side
peak is magnified in the inset.  The diagrams are sample
configurations for region III (bottom wire only), region II (both
wires) and region I (top wire only), from left to right. (c)
Tunneling conductance for region III for several values of
H$_{||}$. } \label{Figure 3}
\end{figure*}

Fig. 3(b) shows the tunneling conductance in regions I, II and III
for sample B with the top and bottom quantum wire split gates set
to a parallel conductance of $\sim 4e^{2}/h$. The resonance at
$\sim$ 2.8 mV is 2D-2D tunneling from the pair of 1 $\mu$m
$\times$ 2 $\mu$m 2D areas shown in Fig. 1. As we move from region
I to region II we observe a doubling of the peak tunneling
conductance.  This is attributed to a doubling of the 2D-2D
tunneling area as we move from only one wire occupied (region I)
to both wires occupied (region II).  The peak tunneling
conductance is halved as we move from region II to region III for
the same reason.  An additional peak is observed at V$_{SD}
\approx$ 13 mV in region I and at V$_{SD} \approx$ - 7 mV in
region III enlarged view in Fig. 3(b). The position of these
additional peaks are sensitively dependent on the top and bottom
split gate voltage and are only visible when one of the wires is
depleted.  It should be noted that these additional peaks are {\em
not} due to second 2D subband tunneling which occurs at $\sim$ 40
mV independent of the top and bottom split gate voltages.  As we
move towards region II the additional peaks move towards and
collapse into the 2D-2D tunneling resonance. For a fixed top and
bottom split gate voltage, the peak position is independent of
magnetic field, but the amplitude can have a strong dependence as
shown in Fig. 3(c).

In region II of Fig. 2 we have observed a 2D-2D tunneling
signature as shown in Fig. 3(b). However, we have not observed a
clear 1D-1D signal as expected.  From previous work on vertically
coupled quantum wires \cite{Tserkovnyak2002} we expect a 1D-1D
conductance signal on the order of $\sim$ 0.01 $\mu$S, which
should be clearly observable in this experiment. Comparing 2D-2D
tunneling results to the tunneling signal from region II we
estimate that any 1D-1D signal buried in the 2D-2D background to
be no larger than $\sim$ 0.001 $\mu$S.

Finally we turn our attention to the side peaks in the tunneling
when only one wire is occupied (regions I and III). From the
strong magnetic field dependence of the amplitude (Fig. 3c) and
the fact that the peak position in $V_{SD}$ depends on the exact
combination of top and bottom split gate voltages, it is tempting
to conclude that the peak arises from 1D-2D tunneling. This
tunneling could be from either the occupied or unoccupied wire to
the 2D region in the other layer.  If these features are indeed
1D-2D tunneling signatures, the tunneling process is quite
different from the resonant 2D-2D tunneling. First, the
insensitivity of peak position with $H_{||}$ suggests an inelastic
origin.  Second, we might expect each of the multiple 1D subbands
to contribute additional tunneling channels; however, no other
tunneling peaks are observed until the second 2D subband tunneling
at $\sim$ 40 mV.

It is important to note the large $V_{SD}$ values of the
additional tunneling structures.  Clearly nonlinear processes can
be important.  One example is lateral transport through the
unoccupied wire at high bias.  This could allow the floating 2D
paddle (see diagrams in Fig. 3b) to contribute to tunneling.
Finally, the effect of a depleted quantum wire at high bias is
unknown in this system. While the origin of the additional
tunneling peaks remains unknown, possibilities of 1D-2D tunneling
and nonlinear transport are being explored to solve this important
problem.

In conclusion, we have fabricated a nanoelectronic structure
consisting of a pair of vertically coupled quantum wires and
measured tunneling between the layers.  The key to fabricating
this device is the combination of EBASE thinning, electron beam
lithography and depletion gates for separate contact.  Transport
measurements demonstrate external control over the number of
occupied subbands in each 1D wire.  With independent contacts,
tunneling is measured between the quantum wire systems.  An easily
identified 2D-2D component is visible, and additional structure is
present at high bias voltages when only one wire is occupied.  The
additional peak in the tunneling measurement is possibly related
to 1D-2D tunneling or nonlinear transport of the unoccupied 1D
wire.  The successful combination of semiconductor
heterostructures and dual-side electron beam lithography can be
used to create a wide variety of electrically coupled
nanostuctures for both fundamental and applied studies.

We thank S. K. Lyo for useful discussions and we acknowledge the
outstanding technical assistance of R. Dunn and D. Tibbets.  This
work has been supported by the Division of Materials Sciences and
Engineering, Office of Basic Energy Sciences, U.S. Department of
Energy. Sandia is a multiprogram laboratory operated by Sandia
Corporation, a Lockheed Martin Company, for the United States
Department of Energy under contract DE-AC04-94AL85000.


\begin{thebibliography}{}

\bibitem{Zulicke2002} U. Zulicke, Science {\bf295}, 810 (2002).

\bibitem{Eugster1991} C. C. Eugster and J. A. del Alamo, Phys. Rev. Lett. {\bf67}, 3586 (1991);

\bibitem{Eugster1994} C. C. Eugster,  {\em et al.}, Appl. Phys. Lett. {\bf64}, 3157 (1994);

\bibitem{Smoliner1996} J. Smoliner, Semicond. Sci. Technol. {\bf11}, 1 (1996).

\bibitem{Ploner2000} G. Ploner, {\em et al.}, Appl. Phys. Lett. {\bf74}, 1758 (1999);

\bibitem{Auslaender2002} O. M. Auslander, {\em et al.}, Science {\bf295}, 825 (2002).

\bibitem{Tserkovnyak2002} Y. Tserkovnyak, {\em et al.}, Phys. Rev. Lett. {\bf89}, 136805
(2002); Phys. Rev. B {\bf68}, 125312 (2003).

\bibitem{Bird2003} T. Morimoto, {\em et al.}, Appl. Phys. Lett. {\bf82}, 3952 (2003).

\bibitem{Governale2002} M. Governale, {\em et al.}, Phys. Rev. B {\bf65}, R140403 (2002).

\bibitem{Bertoni2000} A. Bertoni, {\em et al.}, Phys. Rev. Lett. {\bf84}, 5912 (2000).

\bibitem{EBASE1996} M. V. Weckwerth, {\em et al.}, Supperlatt. Microstruct. {\bf20}, 561 (1996).

\bibitem{Eisenstein1991} J. P. Eisenstein, {\em et al.}, Phys. Rev. B {\bf44}, 6511 (1991).

\bibitem{Smoliner1989} J. Smoliner, {\em et al.}, Phys. Rev. Lett. {\bf63}, 2116 (1989).

\bibitem{Smoliner1995} G. Rainer, {\em et al.}, Phys. Rev. B {\bf51}, 17642 (1995).

\end{thebibliography}
\end{document}